\documentclass{aa}
\usepackage{graphics}
\usepackage{epsfig}

\begin{document}

   \title{The Draco dwarf galaxy in the near-infrared}

   \author{M.-R.L. Cioni\inst{1} \and H.J. Habing\inst{2}}

   \offprints{mrc@roe.ac.uk}

   \institute{SUPA, School of Physics, 
	University of Edinburgh, IfA, 
	Blackford Hill, Edinburgh EH9 3HJ, UK
	\and
	Sterrewacht Leiden, Niels Bohrweg 2, 
	2333 RA Leiden, The Netherlands}

   \date{Received 4 May 2005 / Accepted 11 July 2005}

   \titlerunning{Draco in the near-infrared}

   \abstract{With the  William Herschel Telescope in La  Palma we made
   $IJK_{\mathrm    s}$   observations   of    an   area    of   about
   $40^{\prime}\times30^{\prime}$  of the  Local  Group galaxy  Draco.
   This  allows us  to describe  Draco's late-type  stellar population
   across the whole galaxy at  a photometric level $2$ mag deeper than
   the 2MASS  survey.  We detected  the red giant branch (RGB) and measured
   the magnitude of the tip of  the RGB in the three bands.  From that
   in  the  $I$  band  we  obtain  a  distance  modulus  of  $(m-M)_0=
   19.49\pm0.06{(\mathrm  stat)}\pm0.15{(\mathrm sys)}$,  in excellent
   agreement with a measurement from  RR Lyrae stars.  The peak of the
   $(J-K_{\mathrm  s})_0$ histogram  at different  $M_{K_{\mathrm s}}$
   suggests  that Draco  has a  mean [Fe/H]$=-1.95\pm1.26$  while
   fiducial RGB  tracks of Galactic globular clusters  indicate a mean
   [Fe/H]$=-1.33\pm0.72$  where the error  corresponds to  the spread
   around the  mean value.  There are  significant differences between
   the  colour-magnitude diagrams  of stars  in the  inner,  medium and
   outer   areas   of  the   galaxy.    A   metal  poor   (Z$=0.0004$)
   intermediate-age  population  (about  $1.6$  Gyr  old)  is  clearly
   present  and   emerges  in  particular   between  $6^{\prime}$  and
   $12^{\prime}$  from the  centre  of the  galaxy.  A few  additional
   carbon  star  candidates  have  been  identified  from  both  their
   location in  the colour-magnitude diagram and from  an indication of
   variability.  The large  scale distribution  of late-type  stars is
   smooth  but irregular  in  shape;  this points  at  a variation  of
   inclination with radius.}

   \maketitle

\section{Introduction}
The faint  low surface brightness elliptical Local  Group galaxy Draco
($\alpha_{2000}   =  17^h20^m18.1^s$,   $\delta_{2000}   =  57^{\circ}
55^{\prime}13^{\prime\prime}$; Piatek  et al.  \cite{pia})  is located
at about $80\pm7$ kpc  (Aparicio et al.  \cite{apa}, Nemec \cite{nem})
from the Sun. Bonanos et  al.  (\cite{bon}) derived a distance modulus
of   $(m-M)_0=  19.40\pm   0.02$(stat)$\pm0.15$(syst).  Its   size  is
approximately $35.5^{\prime} \times  24.5^{\prime}$ and about $14$ kpc
along the line  of sight (Aparicio et al.  \cite{apa}). The foreground
reddening ($E(B-V)=0.03\pm0.01$; Stetson  \cite{ste79b}) is lower than
expected  for a  galaxy at  a galactic  latitude  $b=35^{\circ}$.  The
spread in  magnitude among RR Lyrae  stars within the  galaxy shows no
evidence for internal dust (Bonanos et al. \cite{bon}).

The  optical colour-magnitude diagrams  show a  well-defined horizontal
branch (HB)  and a red giant  branch (RGB) without a  signature of the
RGB  bump  although  deep  enough  observations  have  been  performed
(Bellazzini et al.  \cite{bel}). The diagrams are similar  to those of
globular clusters and indicate the  dominance of an old and metal-poor
population  (Stetson \cite{ste79a},  Baade \&  Swope  \cite{bad}). The
location and  width of the RGB  is consistent with  a mean metallicity
[Fe/H]$=-2.0\pm0.8$ dex (Carney \& Seitzer \cite{car}). Differences in
temperature and  spectral properties on  the upper giant  branch stars
indicate   the  presence   of  both   RGB  and   AGB   stars  (Stetson
\cite{ste80}).  The main-sequence  turnoff  corresponds to  an age  of
about $18$  Gyr and there  is an almost negligible  younger population
(Stetson  et al.   \cite{ste85}). Observations  with the  Hubble Space
Telescope  (Grillmair et al.  \cite{gri}) showed  that: (i)  the stars
formed primarily in a single  epoch; (ii) the initial mass function is
very similar to  that in the solar neighbourhood; and  (iii) there is a
non-negligible  fraction  of   blue  stragglers  (Carney  \&  Seitzer,
\cite{car}).  From simulations  of the  colour-magnitude  diagram while
accounting  for  the abundance  pattern  within  the  galaxy Ikuta  \&
Arimoto (\cite{iku})  concluded that over  a long star  forming period
($>3.9-6.5$ Gyr) the  star formation rate was low  ($1-5$\% of that of
the  solar neighbourhood);  the stripping  of interstellar  gas  by the
Milky  Way Galaxy  perhaps  ended  the star  formation  in Draco.  The
intensity of diffuse H$_\alpha$ (Gallagher et al.  \cite{gal}) implies
a mass of the ionised gas below $10$\% of the stellar mass. This limit
is  10 times  larger than  the  upper limit  for the  mass of  neutral
hydrogen (Young \cite{you}).

A  few  hundred  variable   stars  have  been  identified  in  various
surveys.  The most  numerous are  RR Lyrae  stars but  there  are also
Cepheids (Baade  \& Swope  \cite{bad}, Deupreee \&  Hodson \cite{deu},
Nemec \cite{nem},  Bonanos et al. \cite{bon}).  The  Sloan Digital Sky
Survey covered the  entire galaxy; cross-identifications with existing
optical  catalogues  gave $142$  additional  candidate variable  stars
(Rave et al. \cite{rav}).

Star  counts  show  that  the  Draco  galaxy  is  elliptical  with  an
eccentricity $\epsilon  = 0.29\pm0.02$, a position angle  of the major
axis of $88^{\circ}\pm3^{\circ}$ and a  cutoff radius in the long axis
around   $26^{\prime}\pm2^{\prime}$   (Hodge   \cite{hod},  Irwin   \&
Hatzidimitriou  \cite{irw},  Odenkirchen   et  al.   \cite{ode}).  The
Odenkirchen  et al.  paper  shows that  the  spatial distributions  of
giants, horizontal branch, and sub-giant stars down to i$=21.7$ within
an  area  of  $27^{\circ}$  around  the centre  argues  against  tidal
effects.    This   conclusion   was   also  reached   by   Piatek   et
al.(\cite{pia}).   Deeper   optical   photometry   by   Wilkinson   et
al. (\cite{wil})  confirmed a break  in the light profile  at $\approx
25^{\prime}$  and  a  decline   in  the  velocity  dispersion  outside
$30^{\prime}$  which  imply  the  existence of  a  kinematically  cold
population in the outer part of  the galaxy.  On the other hand, there
is no difference  between the scale length of  the distribution of old
($>9$ Gyr) and intermediate-age ($2-3$ Gyr) stars and both populations
were either formed under the same kinematic conditions, or any initial
difference was afterwards erased (Aparicio et al.  \cite{apa}).

Spectra of  giant stars  in Draco have  been obtained since  the early
1970s in  order to  derive the abundance  of the heavier  elements and
other  stellar  parameters  (Hartwick  \&  McClure  \cite{hart}).  The
results showed that  within the galaxy there is a  spread in [Fe/H] of
at  least  1.53  dex  (Shetrone  et  al.  \cite{she01b},  Shetrone  et
al.  \cite{she98}, Zinn  \cite{zin78}) and  that giants  in  Draco are
metal  poor ([Fe/H]$\approx  -2.0$) compared  to giants  in  Milky Way
globular clusters (Canterna  \cite{can}, Stetson \cite{ste84}, Winnick
\cite{win}). In fact the location of stars redward of the fiducial RGB
sequence,  indicate that  these stars  have a  higher  metallicity but
still below [Fe/H]$=-1.45$ dex; they must be carbon stars (Shetrone et
al. \cite{she01b}). Inhomogeneities in the [Ca/H] and [Mg/H] abundance
also  support  an abundance  spread  (Winnick  \cite{win}, Lehnert  et
al. \cite{len}).  Supernova activity may  account for this  spread and
for the  removal of  about $90$\% of  the original gas  content (Smith
\cite{smi}). A handful of carbon  stars were identified by Aaronson et
al. (\cite{aar}) and by Azzopardi et al. (\cite{azz}). One of those is
a symbiotic binary (Munari \cite{mun}) and another is lithium-rich and
probably  a low-mass  AGB star  of  low metallicity  in its  thermally
pulsing phase  after a few third dredge-up  episodes (Dom\'{i}nguez et
al.  \cite{dom}).  Olszewski  et  al. (\cite{ols95})  confirmed  three
carbon stars as velocity variables  which implies that they are likely
binary stars similar to CH stars in the Milky Way halo.  The sample of
confirmed  carbon   stars  was  increased   to  six  by   Shetrone  et
al. (\cite{she01b}).

The  mean velocity  of  the Draco  galaxy,  determined from  different
samples of  stars, is $-293.3\pm1.0$  km/s with a dispersion  of about
$8.5\pm0.7$  km/s   (Hargreaves  et  al.   \cite{har},  Armandroff  et
al.  \cite{arm}). This  dispersion  is significantly  larger than  the
velocity dispersion of  other galaxies in the Local  Group, except for
the Ursa  Minor and  Carina galaxies. The  measured dispersion  is not
dominated by binary  stars (Olszewski et al. \cite{ols})  and the mass
to light ratio ($145^{+116}_{-71}$) is consistent with the presence of
a large quantity of dark matter (Hargreaves et al. \cite{har}).

\begin{figure}
\resizebox{\hsize}{!}{\includegraphics{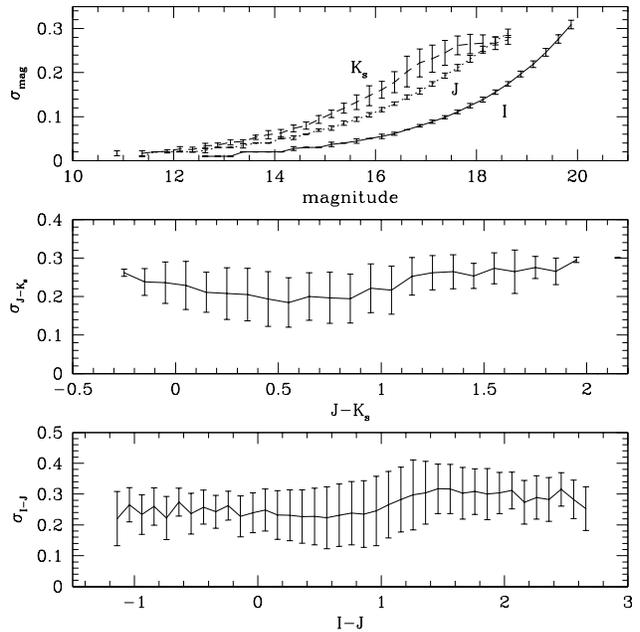}}
\caption{Distribution  of the  mean-error as  a function  of magnitude
({\bf top}) and  colour ($J-K_{\mathrm s}$ - {\bf  middle} and $I-J$ -
{\bf bottom}) for sources in Table 1. Error bars indicate the standard
deviation of  the mean,  they have  been both obtained  with a  bin of
$0.25$  mag in $IJK_{\mathrm  s}$ and  of $0.1$  mag in  colour. Lines
connect  the mean  points of  each bin:  continuous ($I$  and colour),
dotted ($J$) and dashed ($K_{\mathrm s}$).}
\label{err}
\end{figure}

Our present work concerns  the intermediate and old stellar population
in  Draco  and  uses new  mosaic  observations  in  the $I$,  $J$  and
$K_{\mathrm s}$  bands. Observations and data  reduction are described
in Section 2 while in section 3 we discuss the distribution of sources
in the colour-colour and colour-magnitude diagrams at different distances
from  the centre;  we  also discuss  the  contamination by  foreground
stars.  Section 4  presents a  determination  of the  distance to  the
galaxy, a  discussion about variations  in age and metallicity  of the
stellar population across the galaxy  and a discussion of known carbon
stars.  Conclusions are given in Section 5.

\section{Observations \& Data Reduction}
Observations in the  $I$, $J$ and $K_{\mathrm s}$  bands were obtained
with the  4.2 m William Herschel  Telescope (WHT) on  La Palma (Spain)
during the same runs described  in Cioni \& Habing (\cite{cio05}). Due
to the large extent in the sky  a mosaic of images was required in all
three  bands.   In  particular  we  obtained   $11\times9$  images  of
$4.06^{\prime}\times  4.06^{\prime}$ in  the $J$  and  $K_{\mathrm s}$
band  and $6$  images of  $16.2$\arcmin $\times  16.2$\arcmin~  in the
$I$-band. In the $I$-band we  used exposures of $6$s, $60$s and $300$s
and  $J$ and  $K_{\mathrm s}$  images were  obtained by  co-adding $4$
images out of $5$ of a  single dithered position each with an exposure
time of $1.5$s. In total we covered an area of about $40^\prime \times
30^\prime$  centred  at  $\alpha   =  17$:$20$:$12.4$  and  $\delta  =
+57$:$54$:$55$.

The  processing of raw  images was  done using  the IRAF  software and
following   the  same   procedure   described  in   Cioni  \&   Habing
(\cite{cio05})  with  a  few  minor  differences: sky  frames  in  the
near-infrared (near-IR)  wave bands were  obtained averaging uncrowded
target images;  - the same fringe  image was used to  correct long and
short $I$-band exposures. Near-IR images were combined in $11$ columns
per band ($9$  images per column) while $I$-band  images were combined
in one  single image using  the spatial overlap.  Each  near-IR column
was matched  to the $I$-band  image to find source  counterparts. Note
that there  was no  overlap among near-IR  columns ($=$ no  overlap in
right  ascension), while the  overlap in  declination was  quite large
($1^{\prime}$) to compensate for the effects of image distortions.

Sources were extracted using  the SExtractor program (Bertin \& Arnout
\cite{sex}).   The flux  corresponding to  each source  was calculated
within an aperture  of $5$ pixels, a detection  threshold of $2.0$, an
analysis threshold of $1.5$.  In the SExtractor configuration files we
used these parameters as well  as the parameters that characterise the
detector and default values  for the remaining keywords.  Sources were
extracted first in each wave band separately. Afterwards $J$ and
$K_{\mathrm s}$ detections were matched using an association radius of
$6$  pixels   ($6$  pix  $\times   0.238^{\prime\prime}$/pix  $\approx
1.4^{\prime\prime}$)  to  account   for  possible  field  distortions,
although $J$ and $K_{\mathrm s}$ images were aligned before extracting
individual sources.  The nearest source  was kept as a counterpart and
the  resulting catalogue  was  filtered to  retain  only sources  with
$K_{\mathrm s}<-(J-K_{\mathrm s})+18.5$  (this artificial cut is shown
at  faint   magnitudes  and  red  colours  in   Fig.  \ref{jkk}.   The
cross--identification  of $I$--band  sources was  made  separately for
each  near-IR  column.  In   practice  an  $I$-band  column-image  was
extracted from the full $I$-band image to match a given near-IR column
(i.e. by aligning common sources and trimming the area).  Sources were
matched  using the  same criteria  but an  association radius  of $80$
pixels.   Different association  radius  were initially  used and  the
resulting matches  where checked against known  sources. Finally wrong
associations were removed by  accepting only pairs for which $14<I<20$
and $-1<(I-J)<3$.  This criteria combines both position, magnitude and
colour  which guarantees a  statistical match  of similar  sources and
accounts for field  distortions. The $I$-band field of  view is rather
large ($16.2^{\prime}\times 16.2^{\prime}$) and distortions at the far
edges are non-negligible. With a smaller matching radius we would have
lost  a few bright  counterparts (those  located at  the edges  of the
field-of-view). However because of  the image alignment and because we
retained the closest match  within a satisfactory colour and magnitude
range  good results  were  obtained. A  few  mis-matches, between  the
near-IR and $I$-band magnitudes might be present among the faint stars
in  the  catalogue.  For  these  sources the  match  between  $J$  and
$K_{\mathrm  s}$  is  more  reliable  compared to  the  match  between
$JK_{\mathrm s}$ and $I$.

The photometric calibration was performed for each observing night and
it  is  described  in   detail  by  Cioni  \&  Habing  (\cite{cio05}).
Photometric zero-points in the $J$ and $K_{\mathrm s}$ bands have been
corrected for systematic  shifts after cross-identification with 2MASS
data.    This   process   showed   a    systematic   shift   of
$5.25^{\prime\prime}\pm1.75^{\prime\prime}$ between  the astrometry of
the two  data sets were the  error-bar indicates the  dispersion of the
mean. Only  sources within $1.75^{\prime\prime}$ of  this value ($248$
sources)  have been  used to  compare the  photometry  obtained within
similar filters. We obtain:

$J=J^{\mathrm 2MASS}+0.30\pm0.05$

$K_{\mathrm s}=K_{\mathrm s}^{\mathrm 2MASS}+0.30\pm0.07$

 These shifts  (partly due to different filters)  are smaller than
those obtained  by Cioni \& Habing  for the NGC  6822 galaxy, probably
because because  Draco is  not as crowded  as NGC 6822,  although both
galaxies were  observed during poor photometric  conditions (i.e. thin
cirrus).  In  the $I$-band we applied  the same shift  as derived for
NGC 6822 namely $I=I^{\mathrm  DENIS}-0.1 \pm 0.01$. This is justified
because the $I$-band observations of  Draco and NGC 6822 were obtained
during the same night under similar photometric conditions whereas our
near-IR observations were spread over different nights.

\subsection{Catalogue}
Table \ref{table1} shows  the first ten lines of  the full table ({\it
table.dat}) that is  only available in the electronic  edition of this
article. The table contains $2570$  stars detected in all three bands,
$I, J$ and  $K_{\mathrm s}$: columns $1$ and  $2$ list Right Ascension
and Declination  in degrees at the  epoch J2000, columns  $3$, $4$ and
$5$ list  $I$ magnitude, photometric  error and SExtractor  flag ($1=$
Bright  neighbours or  bad pixels  affecting more  than $10$\%  of the
integrated area, $2=$ The object was originally blended with another),
respectively; columns  $6-8$ and  $9-11$ contain the  same information
for the $J$ and $K_s$  bands, respectively. The table lacks sources in
the gap between pairs of $I$--band chips. We include only sources with
SExtractor flag $<4$  in all three wave bands.    Figure \ref{err}
shows the behaviour of photometric errors for all extracted sources as
a function of  magnitude in each wave band and as  a function of $I-J$
and $J-K_{\mathrm s}$ colour.

\begin{figure}
\resizebox{\hsize}{!}{\includegraphics{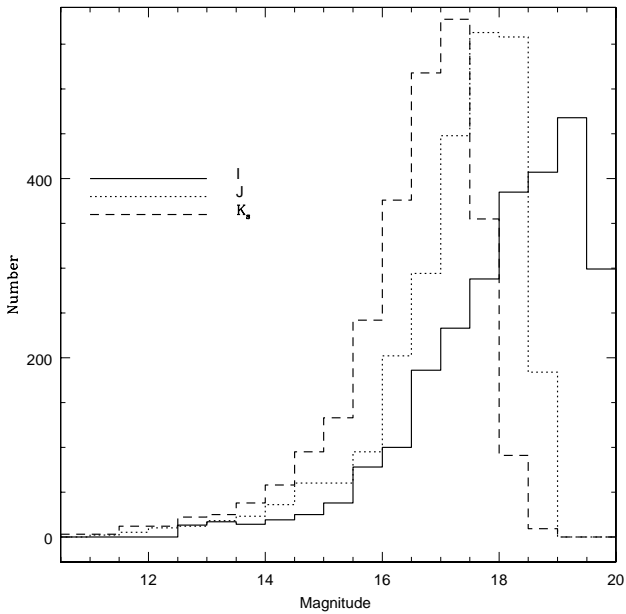}}
\caption{Histogram of  the number of  sources matched among  the three
wave  bands as  a function  of magnitude:  $I$ (continuous  line), $J$
(dotted line)  and $K_{\mathrm  s}$ (dashed line).  Each bin  is $0.5$
mag.}
\label{hist}
\end{figure}

Crowding  is  unimportant  in  Draco  and  there  is  no  issue  about
confusion. The  catalogue is almost complete down  to magnitudes $18$,
$16.8$  and $17.2$  in $I$,$J$  and $K_{\mathrm  s}$ but  we  may have
missed a small percentage of sources because of missing adjacent tiles
of the  mosaic; this effect will  be strongest in  the $I$-band. These
limits correspond to a decrease in the number of sources towards faint
magnitudes  estimated from  Fig. \ref{hist}.  Sources detected  in one
single  band or  only  in the  two  infrared bands  are  a very  small
percentage of those  presented in Table 1; the  catalogue includes all
stars  in the  upper  RGB  magnitudes (Sect.   3).  These single  band
sources or those  detected only in $J$ and $K$  will be made available
upon request to the first author.

\begin{table*}
\caption{Draco catalogue of sources detected in three wave bands.}
\label{table1}
\[
\begin{array}{ccccccccccc}
\hline \hline  \noalign{\smallskip} \alpha &  \delta & I &  \sigma_I &
\mathrm{f}_I  & J &  \sigma_J &  \mathrm{f}_J &  K_s &  \sigma_{K_s} &
\mathrm{f}_{K_s} \\ 260.089081 & 57.691799 &  18.04 & 0.13 & 0 & 17.05
& 0.17 & 0 & 16.16 & 0.07 & 0 \\ 260.090729 & 57.697617 & 19.07 & 0.21
& 0 & 17.05 & 0.17 & 0 &  16.16 & 0.07 & 0 \\ 260.048615 & 57.717018 &
16.65 & 0.07 & 0 & 16.06 & 0.11 & 0 & 15.37 & 0.06 & 0 \\ 260.084015 &
57.717052 & 17.99 & 0.13 & 0 & 16.90  & 0.16 & 0 & 16.08 & 0.07 & 0 \\
260.066681 & 57.717663 & 18.63 & 0.18 & 0 & 18.40 & 0.27 & 0 & 18.09 &
0.08 & 0 \\ 260.086700 & 57.718666 & 17.96 & 0.13 & 0 & 16.90 & 0.16 &
0 & 16.08  & 0.07 & 0 \\ 260.045319  & 57.720440 & 17.92 &  0.13 & 0 &
16.06 & 0.11 & 0 & 15.37 &  0.06 & 0 \\ 260.008881 & 57.723244 & 18.98
& 0.21  & 0 &  18.38 & 0.27  & 0 &  17.14 & 0.08  & 0 \\  260.003357 &
57.725845 & 17.80 & 0.12 & 0 & 18.38  & 0.27 & 0 & 17.14 & 0.08 & 0 \\
260.065857 & 57.726101 & 18.55 & 0.17 & 0 & 18.40 & 0.27 & 0 & 18.09 &
0.08 & 0 \\ \noalign{\smallskip} \hline
\end{array}
\]
\end{table*}

\section{Results}
\subsection{Magnitudes and colours}
\subsubsection{$I$, $J$ and $K_{\mathrm s}$ histograms}
Figure  \ref{hist}  shows  the  histogram  of the  number  of  sources
detected in all three wave bands as a function of magnitude.   The
discontinuity produced by the TRGB  is clearly visible in the $I$-band
but  it  is not  obvious  in the  $J$  and  $K_{\mathrm s}$  magnitude
distributions.  Other discontinuities  may be  present; they  will be
considered in  some detail in Sect.  3.3. The precise  location of the
TRGB,  derived   using  the  same   technique  as  in  Cioni   et  al.
(\cite{cio00}),   is   at   $I=15.54\pm0.06$,   $J=14.89\pm0.10$   and
$K_{\mathrm s}=14.07\pm0.08$.  The algorithm that  determines the TRGB
position starts with an estimated  magnitude and a magnitude range. In
the  $I$  band we  used  $I=15.5$ as  from  the  discontinuity in  the
magnitude distribution  and a range  of $0.5$ mag. This  is consistent
with the value derived by Bellazzini et al.  (\cite{bel}) and accounts
for possible variations between  the two photometric systems. The TRGB
in the  $J$ and $K_{\mathrm  s}$ bands has  not been derived  prior to
this work, and  so we had to derive the  estimate ourselves. Using the
isochrones by Girardi et al. (\cite{gir}) for a $9$ Gyr old metal poor
population  ($Z=0.0004$; this corresponds  to about  [Fe/H]$=-1.7$) we
derived the difference between the  TRGB magnitude in the $I$-band and
those in the  $J$ and $K_{\mathrm s}$ bands; they  are $0.8$ and $1.5$
mag,  respectively. This  results in  initial values  for the  TRGB of
$J=14.7$ and  $K_{\mathrm s}=14.0$  and the same  range of  $0.5$ mag.
Madore \&  Freedman (\cite{mad}) postulated that at  least $100$ stars
in the  upper $1$-magnitude  bin of  the RGB are  needed for  a robust
determination of the TRGB magnitude  (see also Fig.10 in Bellazzini et
al. \cite{bel}) and  we do fulfil this criterium:  we dispose of more
than $200$  stars in the  upper $1$-magnitude bin  in the $I$  and $J$
bands  and  more  than  $100$  in the  $K_{\mathrm  s}$  band.  Errors
associated   to   each  TRGB   determination   are  only   statistical
errors. Systematic errors  due to errors in the  zero-point and in the
calibration  of   the  magnitude  scale   amount  to  $\sigma_I=0.09$,
$\sigma_J=0.11$ and $\sigma_{K_{\mathrm s}}=0.15$. We have not removed
potentially foreground stars;  for Draco this is a  difficult task and
foreground  stars are expected  to have  a continuous  distribution in
magnitude and thus have no effect on the location of the discontinuity
that is  the TRGB.  Further considerations about  foreground stars are
given in Sect.  3.4.

\subsubsection{($I-J$, $I$) colour-magnitude diagram}

\begin{figure}
\resizebox{\hsize}{!}{\includegraphics{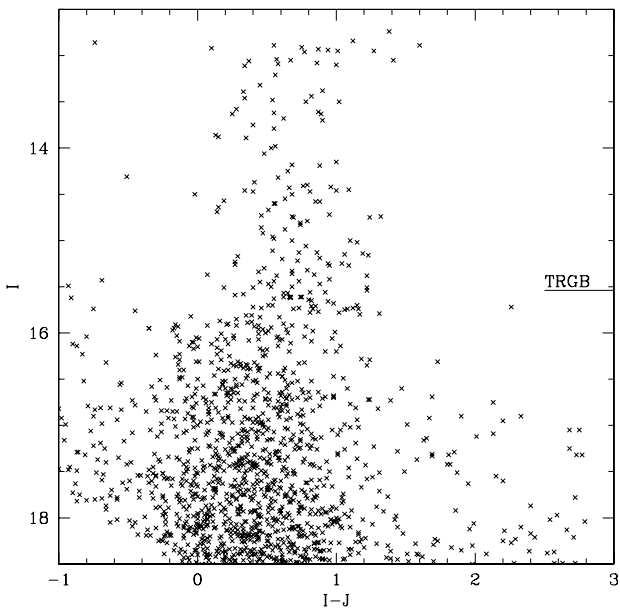}}
\caption{Colour-magnitude  diagram  of stars  detected  in three  wave
bands with $\sigma  _I<0.18$ mag. The location of  the TRGB derived in
this work is indicated.}
\label{iji}
\end{figure}

The   colour  magnitude  diagram   ($I-J$,  $I$)   is  shown   in  Fig.
\ref{iji}. It  is a sparse distribution of  sources and characteristic
features of  such diagrams cannot  easily be traced.  The  location of
the  TRGB marks the  distinction between  fainter sources  that belong
either to  Draco or to the  foreground and brighter  sources that most
probably belong to the foreground except those with $I-J\approx1$ that
may  be AGB  stars  in  Draco.  Note  that  sources with  $I\approx13$
approach the saturation limit  of our observations.  Only sources
with $\sigma_I<0.18$  have been  plotted, this corresponds  to sources
with $I<18.5$ that represent also  the most reliable matches among the
three photometric bands discussed in this paper.

\subsubsection{($I-J$, $J-K_{\mathrm s}$) colour-colour diagram}

\begin{figure}
\resizebox{\hsize}{!}{\includegraphics{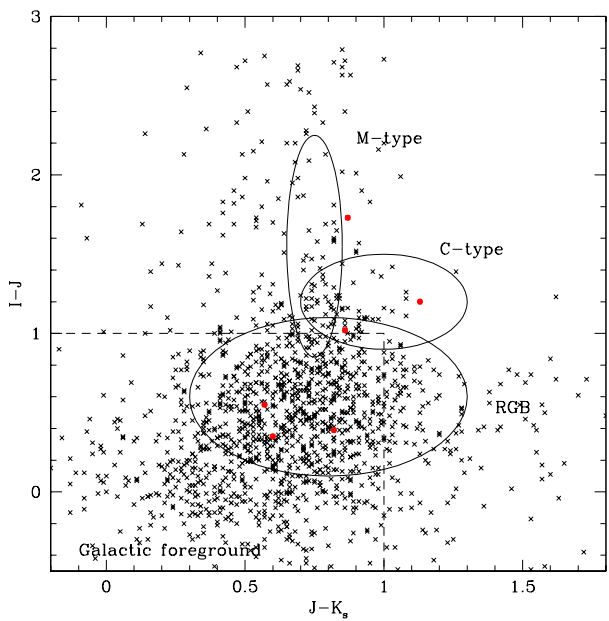}}
\caption{Colour-colour diagram  of stars detected in  three wave bands
with $\sigma_I<0.18$  mag. Regions of expected RGB,  C-type and M-type
AGB stars  as well as  Galactic foreground stars are  indicated. Known
carbon stars are marked with thick dots.}
\label{col}
\end{figure}

The  distribution   of  sources  in  the   colour-colour  diagram  (Fig.
\ref{col})  shows  no  clear-cut   separation  between  sources  of  a
different type  as was  the case for  the Magellanic Clouds  (Cioni et
al. \cite{ciomes})  and for NGC  6822 (Cioni \&  Habing \cite{cio05}).
  As  in  Fig.  \ref{iji}   we  have  plotted  only  sources  with
$\sigma_I<0.18$.  Most  of  the  sources  are  concentrated  at  about
$I-J=0.5$  and  $J-K_{\mathrm s}=0.75$.  There  are  usually very  few
foreground stars  with $J-K_{\mathrm  s}>1$ and $I-J>1$,  although the
expected region of RGB stars  clearly overlaps with that of foreground
stars. Only  two known carbon stars  seem to fall  within the expected
C-type  region,  some  are  much  fainter while  one  is  considerably
brighter  and approaches  perhaps the  region where  M-type  AGB stars
could  be located. It  is important  to remember  that the  colours of
stars of a  different type, and therefore the  expected regions in the
colour-colour diagram, depend on the age and on the metallicity of the
stellar population.

\subsubsection{Near-IR colour-magnitude diagram: confirmed members
and non-members of Draco}

\begin{figure*}
\begin{minipage}[b]{.5\linewidth}
\resizebox{\hsize}{!}{\includegraphics{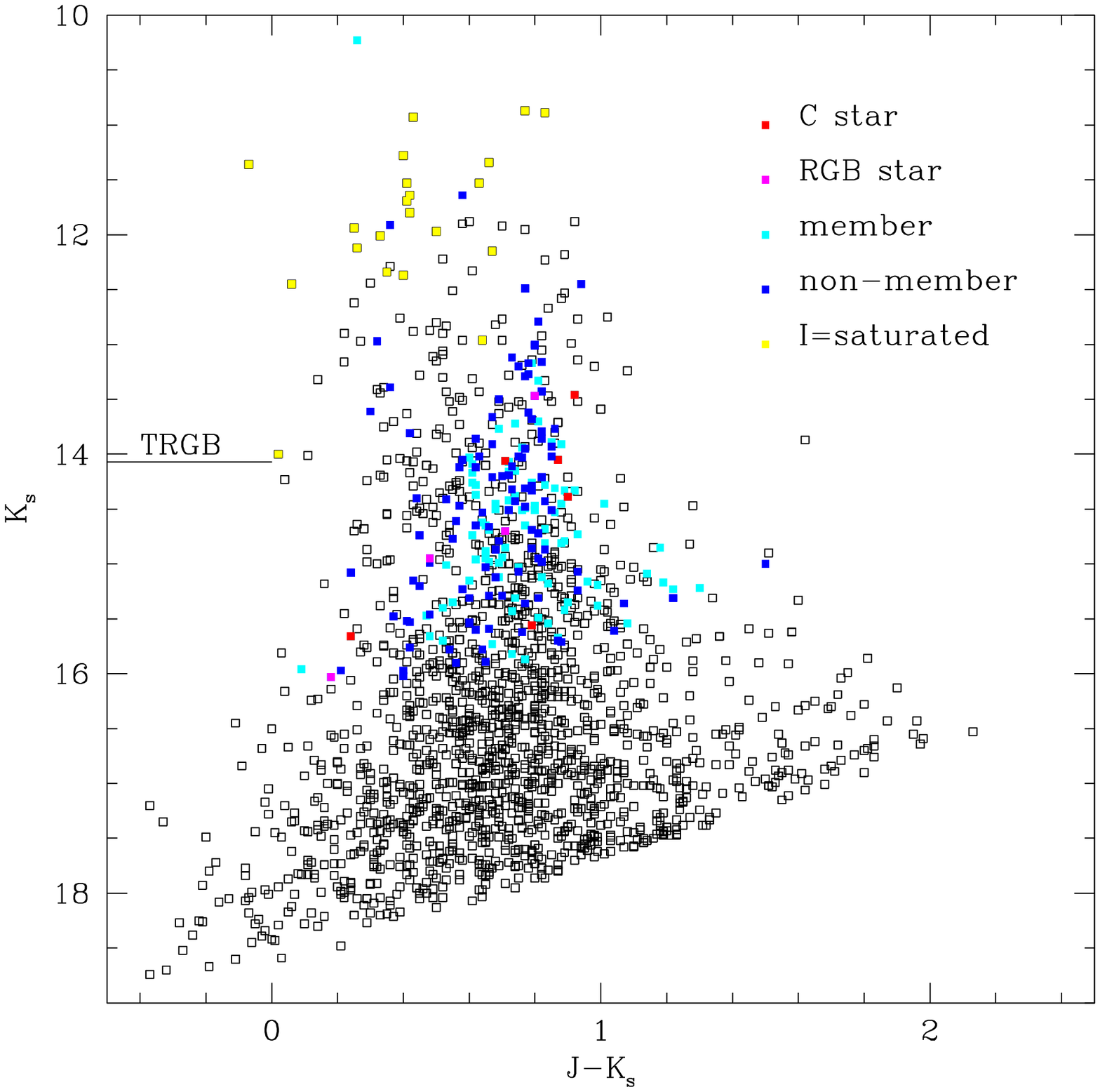}}
\end{minipage}
\begin{minipage}[b]{.5\linewidth}
\resizebox{\hsize}{!}{\includegraphics{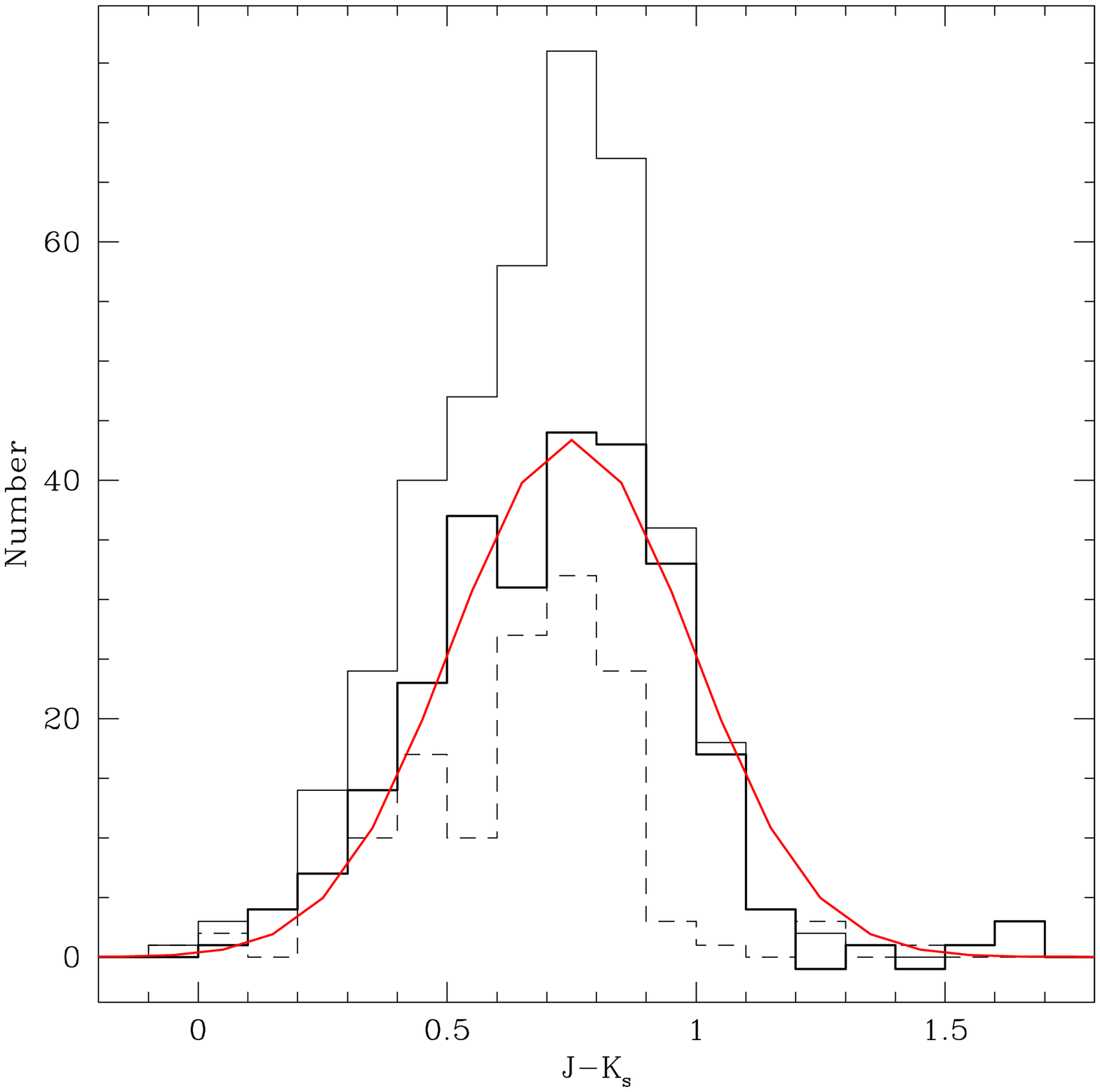}}
\end{minipage}
\caption{({\bf  Left}) Near-IR  colour-magnitude diagram  of  stars in
Draco.   Known   carbon   stars,   known  RGB   stars   (Shetrone   et
al. \cite{she01a}) and  stars that are confirmed not  to be members of
the   galaxy   (Olszewski   et   al.   \cite{ols95},   Armandroff   et
al. \cite{arm})  are indicated  with different colours.  ({\bf Right})
Histogram  of   the  ($J-K_{\mathrm  s}$)  colour   for  sources  with
$K_{\mathrm  s}<15.5$  (thin  continuous line),  confirmed  non-member
sources (dashed  line) and the subtracted  histogram (thick continuous
line).  The  best  fit  Gaussian  has  $\mu=0.74$,  $\sigma=0.24$  and
FWHM$=0.56$. The width of each bin is $0.1$ mag.} 
\label{jkk}
\end{figure*}

The  colour-magnitude diagram ($J-K_{\mathrm  s}$, $K_{\mathrm  s}$) of
Draco stars (Fig. \ref{jkk}) shows  a well defined and broad red giant
branch. To  identify the main  stellar populations in this  diagram we
cross-correlated our catalogue (Table 1) with confirmed carbon and RGB
star  samples and  with non-member  sources  as given  by Shetrone  et
al.  (\cite{she01a}).   We also  cross-correlated  our catalogue  with
foreground dwarfs  identified by  Olszewski et al.  (\cite{ols95}) and
with the catalogue of  confirmed members and non-members by Armandroff
et al.  (\cite{arm}). These cross-identifications have been made using
only coordinates,  except for  the known carbon  stars that  have also
been checked in  the corresponding images, and a  few mismatches might
have   occurred,  especially  among   the  faintest   counterparts.  In
Fig. \ref{jkk} we used colours to mark stars identified by agreement in
position  and  stars that  are  saturated  in  the $I$-band  (possibly
foreground  stars);  the colours  are  better  visible  in the  on-line
version of  the article than  in its printed version.  The sensitivity
limit of previous observations  limited our identifications to sources
with  $K_{\mathrm s}<15$.  Among sources  brighter than  the  TRGB the
brightest magnitudes  are likely foreground stars. Many  of these have
saturated   the    $I$-band   photometry   and    have   predominantly
$(J-K_{\mathrm s})<0.8$. Redder  sources could be intermediate-age AGB
stars. One likely thermally  pulsating carbon star has been identified
in this  region (Sect.   4.2). Along the  main RGB  sequence confirmed
members and non-members are equally  distributed and do not single out
a specific  region in  the diagram.  Figure  \ref{jkk} shows  also the
histogram of the number of sources as a function of colour for the full
sample  (Table 1),  for the  confirmed  foreground stars  and for  the
sample  that   remains  after  subtraction  of   the  foreground.  The
foreground component  dominates at colours  $(J-K_{\mathrm s})<0.8$ and
becomes negligible  at redder colours.  Once the foreground  stars have
been removed  from the full  sample the distribution of  sources looks
fairly symmetric and  can be described by a  Gaussian with $\mu=0.74$,
$\sigma=0.24$.

\subsection{Foreground stars from 2MASS}
Despite the  availability of the  precise location of  some foreground
stars  discussed above  (Armandroff  et al.  \cite{arm}, Olszewski  et
al. \cite{ols}, Shetrone et al. \cite{she01a}), they cannot be used to
estimate  the overall  foreground contamination  because  their parent
sample was not complete. Infact Olszewski et al. (\cite{ols}) observed
the  brightest  and  reddest  objects  selected  from  the  literature
including additional giants near the empirical RGB of Draco.

\begin{figure}
\resizebox{\hsize}{!}{\includegraphics{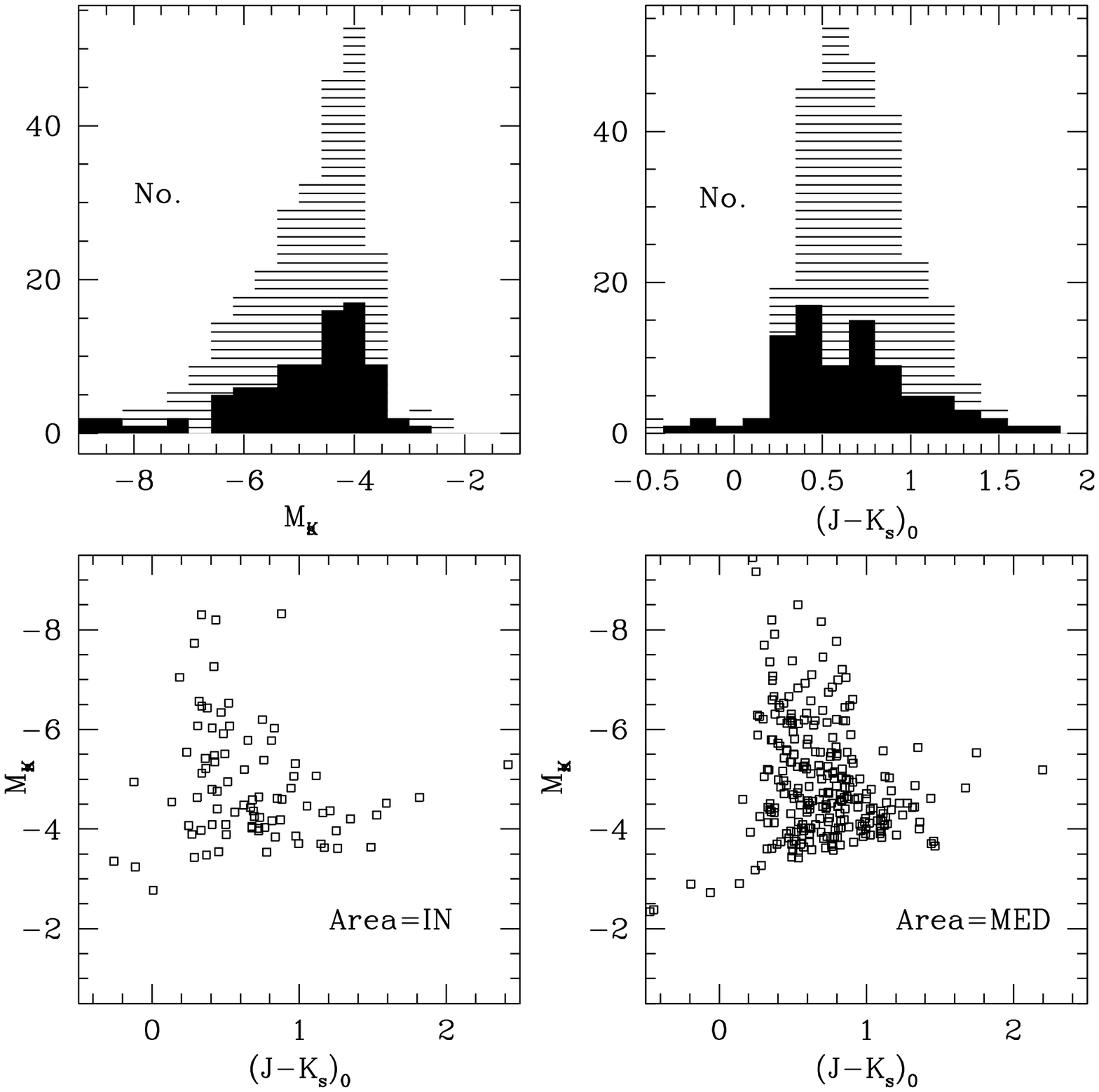}}
\caption{(Upper)  Magnitude  and  colour  distribution  of  foreground
sources  extracted  from  the  2MASS  catalogue  in  the  vicinity  of
Draco. Filled and  shaded histograms represent number counts  in an an
off-set area  equivalent to  the IN and  MED area in  Fig. \ref{cmds},
respectively. The  width of  each bin is  $0.4$ mag  in $M_{K_{\mathrm
s}}$  and  $0.15$  mag  in  $J-K_{\mathrm s}$.   (Lower)  The  near-IR
colour-magnitude  diagram  of  foreground  sources covering  the  same
areas.}
\label{fore}
\end{figure}

We extracted from the 2MASS  catalogue sources within an area of $0.2$
deg$^2$  $0^{\circ}.4$ to  the North  of  the centre  of Draco.   This
region is sufficiently  close to Draco to provide  a comparable sample
of  foreground stars  but  far enough  to  contain very  few, if  any,
genuine  Draco   sources.  Absolute  magnitudes   have  been  assigned
supposing that  they have  the same distance  modulus as the  stars in
Draco and that the absorption is equal to that discussed in Sect. 4.1.
Figure \ref{fore} shows  the distribution of the stars  in the off-set
areas in  magnitude, colour and  in the colour-magnitude  diagram; the
off-set areas were  of two sizes that coincide with  the IN region and
the MED region discussed in  the next Sect. 2MASS observations are not
sensitive enough to a foreground contribution directly comparable with
our  observations  but they  indicate  how  the  foreground stars  are
distributed on the sky.  Foreground stars cover almost homogeneously a
region with $0.2<(J-K_s)_0<0.9$  independently of their $M_{K_{\mathrm
s}}$ mag.  Their number increases  in proportion to the  surface area.
Sources with $(J-K_s)_0>0.9$ are close to the detection limit of 2MASS
and   have   therefore    larger   errors.    A   comparison   between
Fig.  \ref{hist} and Fig.  \ref{fore} shows  no gap  at $M_{K_{\mathrm
s}}\approx-5$ ($K_{\mathrm  s}\approx 14$); this  confirms our earlier
assumption  that one  may find  the  TRGB without  first removing  the
foreground.  The  gap visible in the  colour histogram ($(J-K_{\mathrm
s})\approx0.55$) may have been  caused by differences among the nature
of foreground stars (Cioni  et al. \cite{cio00a}, Nikolaev \& Weinberg
\cite{nik}), see also Fig. \ref{jkk}.

\subsection{Distribution in three concentric areas}
\begin{figure*}
\resizebox{\hsize}{!}{\includegraphics{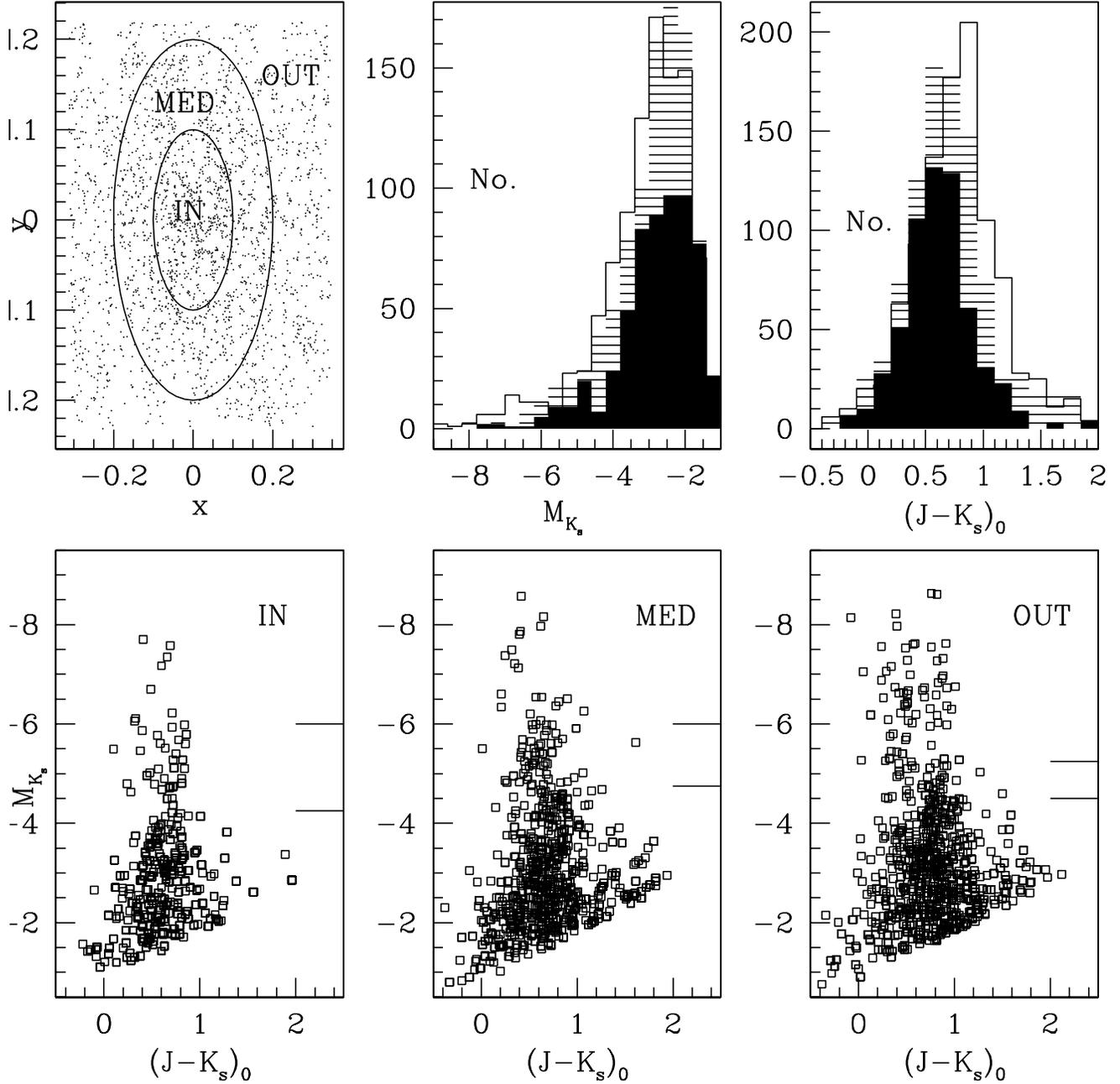}}
\caption{(Upper)  Distribution of Draco  sources in  the plane  of the
sky. Right ascension and declination  have been converted into x and y
coordinates with the centre of  the galaxy (Piatek et al. (\cite{pia})
as origin. We consider three different regions (inner (IN $\rho<0.1$),
medium (MED,  $0.1<\rho<0.2$) and  outer (OUT, $\rho>0.2$).  The three
histograms  of absolute  magnitude $M_{K_s}$  and the  three  of colour
$(J-K_{\mathrm s})$ are shown  with filled, shaded and empty histogram
bins, respectively.  The  width of each bin is  $0.4$ mag in $M_{K_s}$
and  $0.15$   mag  in   $(J-K_{\mathrm  s})$.   (Lower)   The  near-IR
colour-magnitude diagram of sources in  the same three regions, IN, MED
and  OUT.  Horizontal  lines  indicate  the  approximate  position  of
discontinuities in the star counts.}
\label{cmds}
\end{figure*}

We take for the centre of the galaxy the location derived by Piatek et
al. (\cite{pia}).  We have  converted Right Ascension  and Declination
into  circular coordinates  and have  divided the  area  surveyed into
three regions: $\rho<0.1$  (inner=IN), $0.1<\rho<0.2$ (medium=MED) and
$\rho>0.2$ (outer= OUT); see Fig. \ref{cmds}.  Sources are distributed
fairly  homogeneously across  the galaxy  and  do not  cluster in  the
centre.  However, there  is  a non-negligible  difference between  the
near-IR  colour-magnitude diagram  of stars  in three  aspects:  {(i) a
variation in the  expected location of the TRGB},  (ii) a variation in
the number  of stars brighter than  the TRGB and (iii)  a variation in
the colour of the RGB.

 The discontinuity attributed to the TRGB from the distribution of
all  sources  corresponds  to  $M_{K_{\mathrm  s}}=-5.42$  (using  the
distance modulus derived in Sect. 4.1). Although the number of objects
around this magnitude is smaller  within each region it is possible to
distinguish other discontinuities. These are indicated with horizontal
lines in Fig. \ref{cmds}.  Considering for example the discontinuities
at $M_{K_{\mathrm s}}=-4.25$ (IN), $-4.75$ (MED) and $-5.25$ (OUT) one
may conclude that there is a  variation of about $1.0$ mag between the
inner and the outer region in  the sense: the TRGB in the inner region
is fainter than the TRGB in the outer region by about $1.0$ mag.  The
photometric error of each single source is dominated by the systematic
error in the distance modulus,  but as we are interested in variations
in  magnitude  the  only  important  error is  that  in  the  apparent
magnitude; this is  about $0.07$ mag (cfg. Fig.  \ref{err}).  If,
in  the worst-case scenario,  we assume  that all  sources in  the OUT
region of Fig. \ref{cmds} belong  to the foreground, then we find that
about $15$\% and $45$\% of the sources in the IN and MED region should
also belong to the foreground;  most of them would have $M_{K_{\mathrm
s}}>-4$.

The  histogram  of  the  $(J-K_{\mathrm  s})_0$ shows  that  the  peak
increases by about $0.1$  mag by going from IN to MED  to OUT. This is
about equal to  the error in the colour of each  single source, but the
error on the  peak of each histogram decreases  as $1/\sqrt{N}$ and is
of  the  order  of   $0.01-0.02$  mag.     Unless  the  foreground
distribution   strongly   alters  the   shape   of  these   histograms
(cfg. Fig.  \ref{fore}) we may consider  the increase of  $0.1$ mag as
real.

In the colour-magnitude diagram of  the IN region the RGB clearly shows
a  jump in  the number  of sources  around  $M_{K_{\mathrm s}}=-4.25$,
although   the  diagram  clearly   contains  brighter   magnitudes  at
approximately the same colour until $M_{K_{\mathrm s}}=-6$. There are a
few, even brighter  sources but they have bluer  colours and were found
to be the foreground stars discussed  below. In the MED region a major
discontinuity occurs around $M_{K_{\mathrm s}}=-4.75$ as well as a cut
off in  the number of  sources around $M_{K_{\mathrm  s}}=-6$. Sources
with $-4.75>  M_{K_{\mathrm s}}>-6$ trace  a clear branch  about $0.4$
mag wide  and have  bluer colours  compared to the  main RGB  traced by
fainter sources.  Their colour  is also bluer  than stars  with similar
magnitudes  in  the IN  region.  These  sources  could be  metal  poor
intermediate-age AGB  stars. Some  or all of  the very  bright sources
could  be foreground  sources.  In the  OUT  region there  is a  major
discontinuity around  $M_{K_{\mathrm s}}=-5.25$ and  perhaps a fainter
discontinuity  around  $M_{K_{\mathrm   s}}=-4.5$.  At  much  brighter
magnitudes  ($M_{K_{\mathrm s}}<-6$)  sources  are distributed  rather
homogeneously  within  a colour  range  of  about  $1$ mag.  They  most
probably belong to  the foreground, although a few  AGB stars may hide
among them.

\subsection{Spatial distribution}
\begin{figure}
\resizebox{\hsize}{!}{\includegraphics{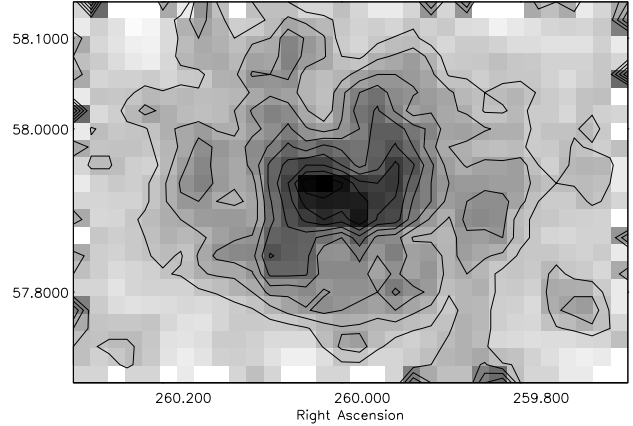}}
\caption{Logarithmic  and  smoothed   density  distribution  of  stars
detected in Draco in $32\times22$ bins of about $1^{\prime}.44$ square
each. Darker regions correspond to a higher density. Contours are from
$3$ to $11$ in steps of $1$.}
\label{map}
\end{figure}

Figure \ref{map} shows the number density of sources in $table.dat$ in
$32\times22$  bins  where  a   single  element  corresponds  to  about
$1^{\prime}.44$  square, which  is equal  to about  $27$ pc.   The map
covers  an area  of about  $40^{\prime}\times30^{\prime}$ and  this is
smaller  than the  cut-off radius  of the  density profile  at optical
wavelengths  ($25^{\prime}-28^{\prime}$)(Hodge  \cite{hod},  Irwin  \&
Hatzidimitriou \cite{irw},  Odenkirchen et al.   \cite{ode}, Wilkinson
et al.  \cite{wil}).  The  source density in  Fig. \ref{map}  has been
smoothed  using  a  box  car  function  of width  $=2$  prior  to  the
construction of  the grey scale  image where higher  concentrations of
sources are  indicated by darker regions.   The distribution, strongly
dominated  by  RGB  stars,  appears  regular and  elliptical  with  an
ellipticity     $\epsilon=0^{\circ}.27$    (where    $\epsilon=1-b/a$;
$a=13^{\prime}.2$ is the semi-major-axis  and $b =9^{\prime}.6$ is the
semi-minor-axis); and $PA  \approx 90^{\circ}$.  Intermediate contours
(at $5$,  $6$ and $7$)  show an enhancement  of the number  of sources
towards  the NW  and the  SE corners,  suggesting a  variation  in the
inclination of the elliptical structure with radius.

\section{Discussion}
\subsection{Determination of the distance to the Galaxy}
The $I$--band  magnitude of  the TRGB depends  only weakly on  age and
metallicity (Salaris \& Cassisi \cite{sal})  and can thus be used as a
standard  candle;   see  Lee  et   al.  (\cite{lee}).  We   adopt  the
interstellar  extinction  as measured  by  Stetson (\cite{ste79b})  of
$E(B-V)=0.03$, and the absorption in  the $I$, $J$ and $K_{\mathrm s}$
band  derived   using  Cardelli's  (\cite{cardi})   law.   These  are:
$A_I=0.05$,  $A_J=0.02$   and  $A_K=0.01$.   Combining   the  apparent
$I$--band TRGB magnitude derived in the previous section with the most
recent absolute calibration of the  TRGB magnitude in the $I$--band by
Bellazzini   et  al.   (\cite{bel}),  $M_I=-4.04\pm0.12$,   we  obtain
$(m-M)_0=19.49\pm0.06  {\mathrm  (stat)}\pm0.15{\mathrm (sys)}$;  this
corresponds to  a distance  of $79 \pm  10$ kpc. The  systematic error
quoted  includes  the error  on  the  absolute  magnitude and  on  the
zero-point   in  the  $I$-band   ($\sigma_I=0.09$;  Cioni   \&  Habing
\cite{cio05})  as  well as  the  error  from  the calibration  of  the
photometry scale ($\sigma_I=0.02$).

Our determination  of the distance  modulus is in very  good agreement
with the  most recent determination of  the distance to  the galaxy by
Bonanos et al.  (\cite{bon}) from  observations of RR Lyrae stars.  It
is  about  $0.3$  magnitudes   smaller  than  the  value  obtained  by
Bellazzini  et al.  (\cite{bel}) who  already noticed  the discrepancy
between  their  result  and   other  measurements  in  the  literature
(e.g.  Mateo  \cite{mat}) and  they  attributed  it  to their  adopted
zero-age    horizontal    branch    distance   scale    (Ferraro    et
al.  \cite{fer99}).  Bellazzini  et al.  determined for  the  TRGB the
value  $I=15.8\pm0.2$ (their Fig.12),  and this  is $0.26$  mag larger
than  ours,  an  acceptable  difference  when  all  uncertainties  are
considered.  It  may be  that  a shift  in  the  photometric scale  is
responsible for this  discrepancy but on the other  hand Bellazzini et
al.  (\cite{bel}) did not have  enough sources to safely determine the
TRGB  position although  they carefully  introduced  artificial stars.
Our results of the distance to the galaxy agree with the determination
by Aparicio  et al.  (\cite{apa})  and by Nemec (\cite{nem})  based on
$B$  and $R$  band  photometry  and observations  of  RR Lyrae  stars,
respectively.

\subsection{Carbon stars and other AGB stars}
\label{carbon}
Shetrone et al. (\cite{she01a}) spectroscopically observed a sample of
stars located on the red side  of the fiducial RGB in the ($B-V$, $V$)
colour-magnitude diagram.  They increased the number  and membership of
known  carbon stars  to $6$  and identified  non-members  by measuring
radial velocities.  The photometry from  our study and from  the 2MASS
counterparts  are indicated in  Table \ref{table2}.  In this  case the
cross-identification between our catalogue and the 2MASS catalogue has
been checked on  the observed images. Source names  are those commonly
used  except for  $C5$ and  $C6$ that  have been  introduced by  us for
completeness.  Four out  of these six carbon stars  are located around
the TRGB. The detection of Li in the brightest of them (Domin\'{i}guez
et  al.   \cite{dom})  suggests  it  is  a   thermally  pulsating  AGB
star. Figure  \ref{jkk} shows  that around the  location of  this star
there  is  a  handful  of   stars  too  red  to  be  foreground  stars
($J-K_{\mathrm   s}>0.8$),   that  could   eventually   also  be   AGB
stars. Spectra are  needed to confirm their nature.  If the population
of Draco  has a large  spread in metallicity  the TRGB will  either be
smeared out  or be  found at different  $K_{\mathrm s}$  magnitudes in
different  locations; it  is  not necessarily  true  that the  fainter
carbon stars  will be below the  TRGB (Sect. 3.3). It  is difficult to
distinguish  oxygen-rich  AGB stars,  if  any  is  present, only  from
photometric   observations    of   a    galaxy   with   a    low   AGB
component. Oxygen-rich AGB stars will be located above the TRGB and at
$J-K_{\mathrm s}\approx0.8$ where foreground stars are also present or
at $J-K_{\mathrm  s}>0.8$, where carbon stars might  be present. There
are too few sources in this region in Fig. \ref{jkk}; below we give an
alternative to find AGB candidates.

\begin{figure*}
\begin{minipage}{0.5\textwidth}
\resizebox{\hsize}{!}{\includegraphics{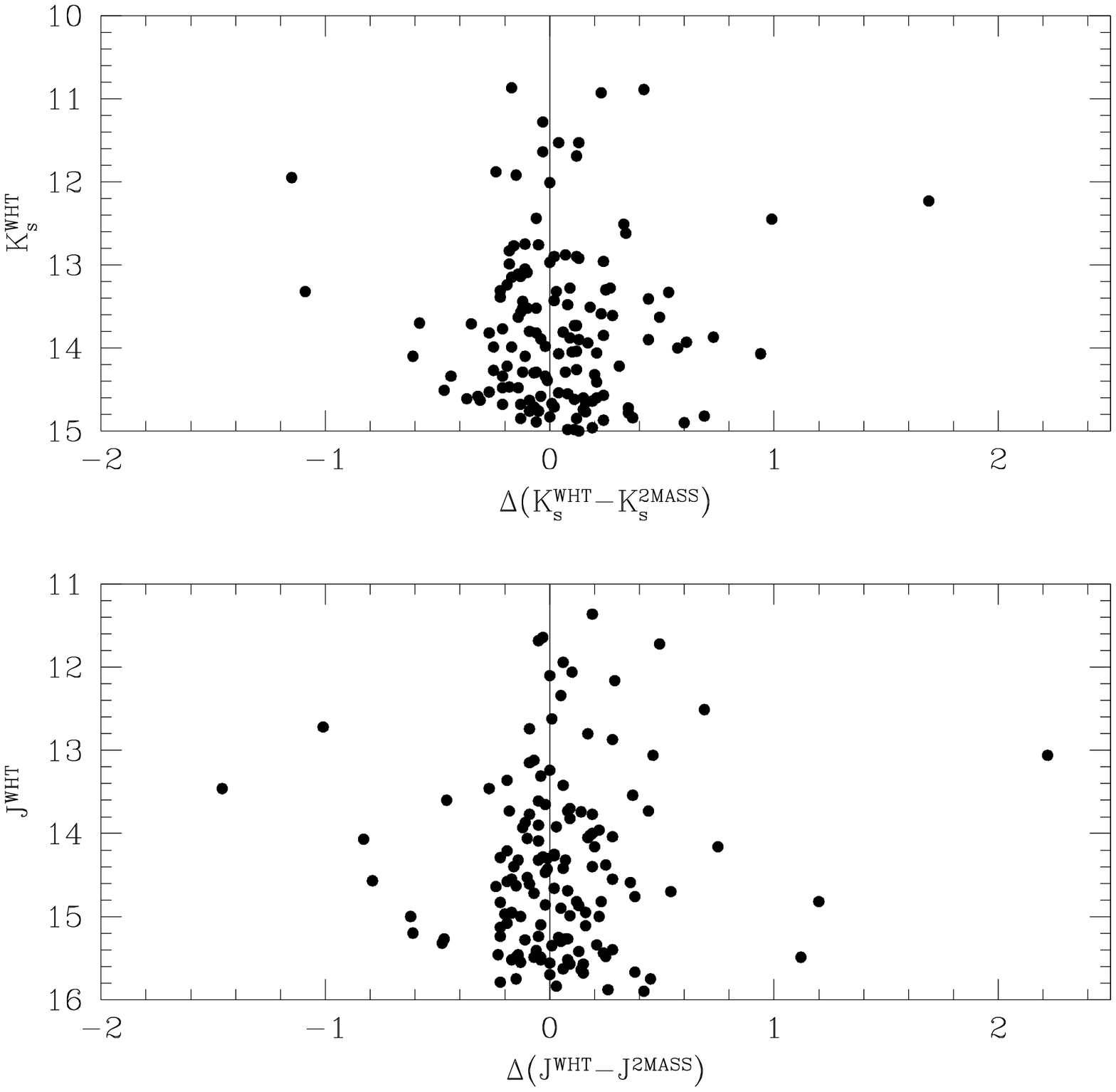}}
\end{minipage}
\hfill
\begin{minipage}{0.5\textwidth}
\resizebox{\hsize}{!}{\includegraphics{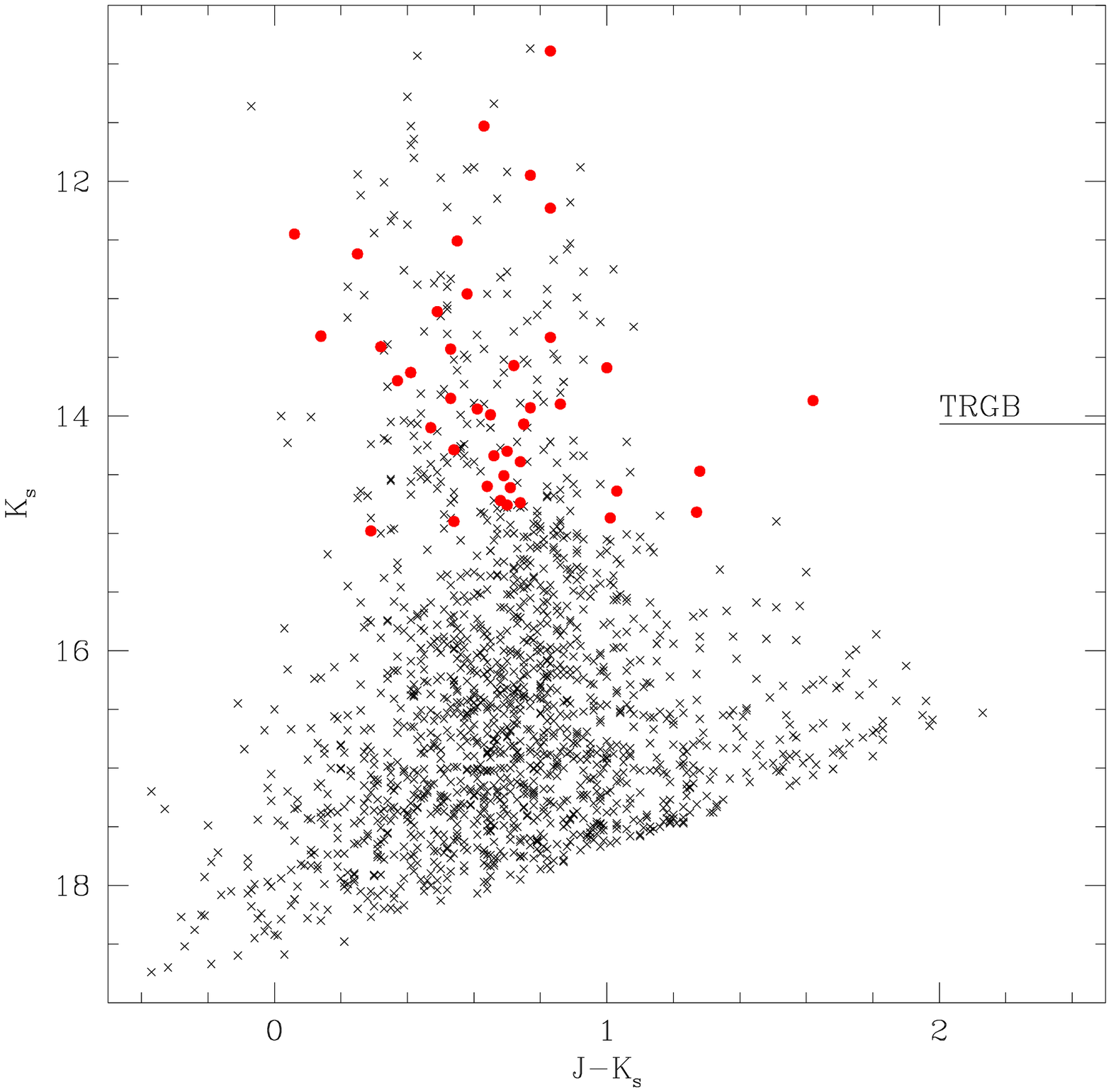}}
\end{minipage}
\caption{({\bf  Left}): Magnitude  difference, $J$  ({\bf  lower}) and
$K_{\mathrm  s}$   ({\bf  upper}),  between   the  near-IR  photometry
presented  in this  article  and  the 2MASS  photometry  for the  same
sources  in  Draco.  Variable  sources  depart from  a  line  at  zero
values.  Note that  that WHT  photometry  has been  adjusted to  2MASS
photometry before  calculating the magnitude difference as  in Sect. 2.
({\bf Right}): Near-IR colour-magnitude diagram of Draco (WHT sources)
where thick  dots indicate  stars with more  then $0.2$  difference in
both the $J$ and $K_{\mathrm s}$ bands.}
\label{var}
\end{figure*}

AGB stars often experience variations in brightness with a long period
and  a  large amplitude.  With  only  two  infrared epochs  we  cannot
determine the periodicity  of a magnitude variation but  we can obtain
an indication  of the  minimum amplitude that  such a  variation would
have.  Figure   \ref{var}  shows   the  difference  between   $J$  and
$K_{\mathrm  s}$ magnitudes  from the  observations described  in this
paper   and   from  the   measurements   extracted   from  the   2MASS
catalogue. The photometry given in Table \ref{table1} has been already
adjusted to  the 2MASS photometry  scale. For a significant  number of
sources $\Delta J>0.2$ mag  and $\Delta K_{\mathrm s}>0.2$), these are
candidate   variable  stars   and  their   position  in   the  near-IR
colour-magnitude  diagram  is  indicated  in  the right  panel  of  the
figure. About  half of the sources differing  in $J$-magnitude between
the two  catalogues are brighter than  the TRGB, up to  $2$ mag, while
the other half  extends to $1$ mag below the  TRGB. These sources with
magnitudes around the  TRGB also have red colours.  They are likely AGB
stars and suitable targets for a spectroscopic study.

\begin{table*}
\caption{Near-IR photometry of known carbon stars}
\label{table2}
\[
\begin{array}{lccccccccccccc}
\hline \hline \noalign{\smallskip} \mathrm{Source}  & I & \sigma_I & J
& \sigma_J  & K_s  & \sigma_{K_s} &  J^{2MASS} &  \sigma_{J^{2MASS}} &
H^{2MASS} & \sigma_{H^{2MASS}}  & K_s^{2MASS} & \sigma_{K_s^{2MASS}} &
2MASS-ID\\ 461 & 16.31 & 0.06 & 14.58  & 0.06 & 13.71 & 0.04 & 14.77 &
0.05 & 14.15 &  0.05 & 14.06 & 0.06 & 17194237+5758376\\  68 & 16.81 &
0.08 &  16.26 & 0.12 & 15.69  & 0.08 & 16.35  & 0.10 & 15.65  & 0.12 &
15.56 & 0.19 & 17195726+5755042\\ 3203 & 15.78 & 0.05 & 14.76 & 0.06 &
13.90  &  0.05 &  14.38  & 0.03  &  13.71  & 0.04  &  13.46  & 0.04  &
17195764+5750054\\ J &  17.75 & 0.12 &  17.41 & 0.20 & 17.35  & 0.08 &
14.92 & 0.04 & 14.22 & 0.05 & 14.05 & 0.06 & 17200068+5753464\\ 3237 &
15.81 & 0.05  & 15.42 & 0.08 & 14.60  & 0.06 & 15.29 &  0.06 & 14.54 &
0.06 & 14.39 & 0.10 & 17203352+5750196\\  578 & 16.49 & 0.07 & 16.14 &
0.12 &  15.54 & 0.07 & 15.90  & 0.10 & 15.49  & 0.14 & 15.66  & 0.29 &
17203883+5759346\\ \noalign{\smallskip} \hline
\end{array}
\]
\end{table*}


\subsection{Variations in age and metallicity}
\subsubsection{The population in the whole area}
\begin{figure*}
\begin{minipage}[b]{.5\linewidth}
\resizebox{\hsize}{!}{\includegraphics{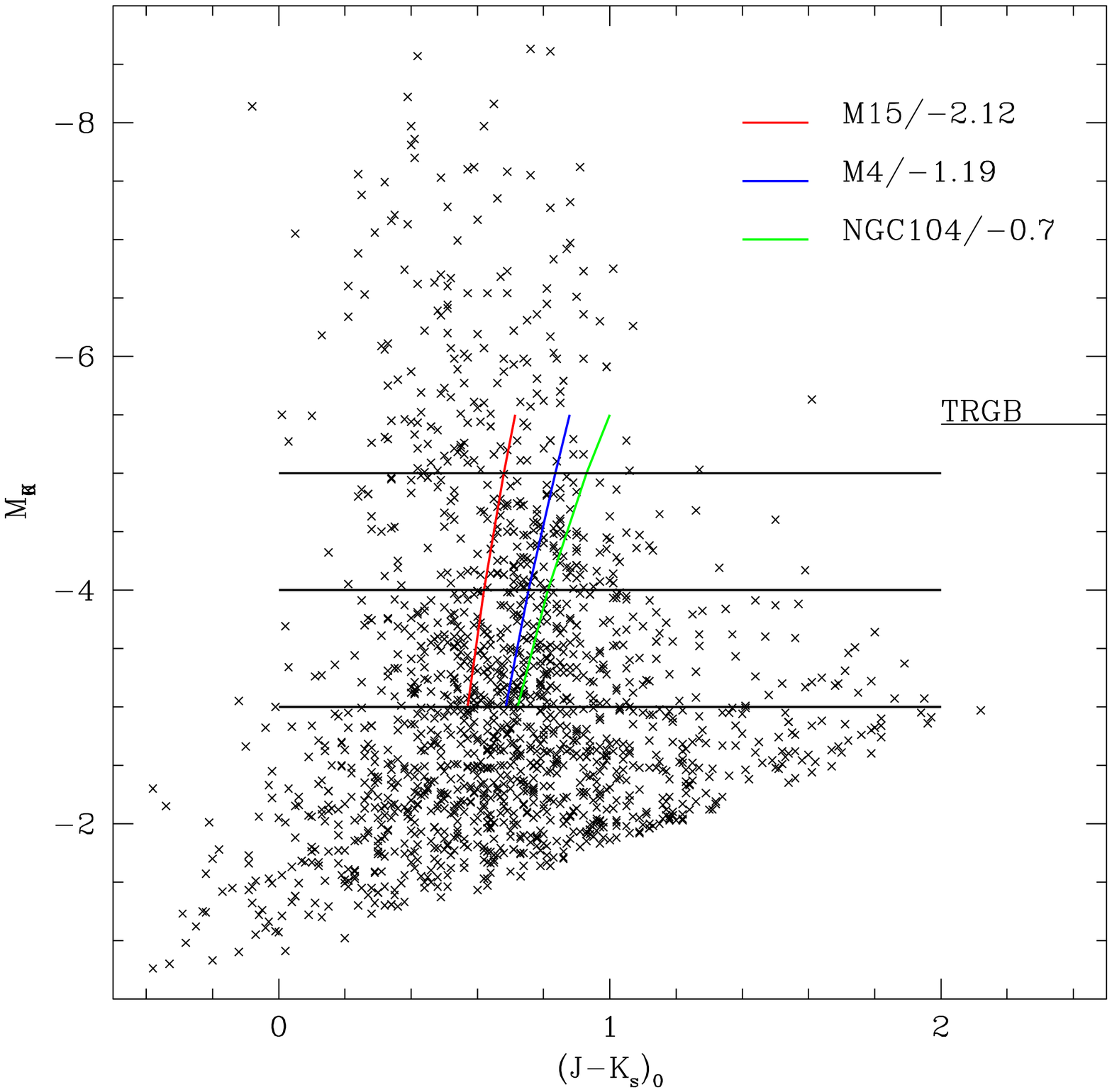}}
\end{minipage}
\begin{minipage}[b]{.5\linewidth}
\resizebox{\hsize}{!}{\includegraphics{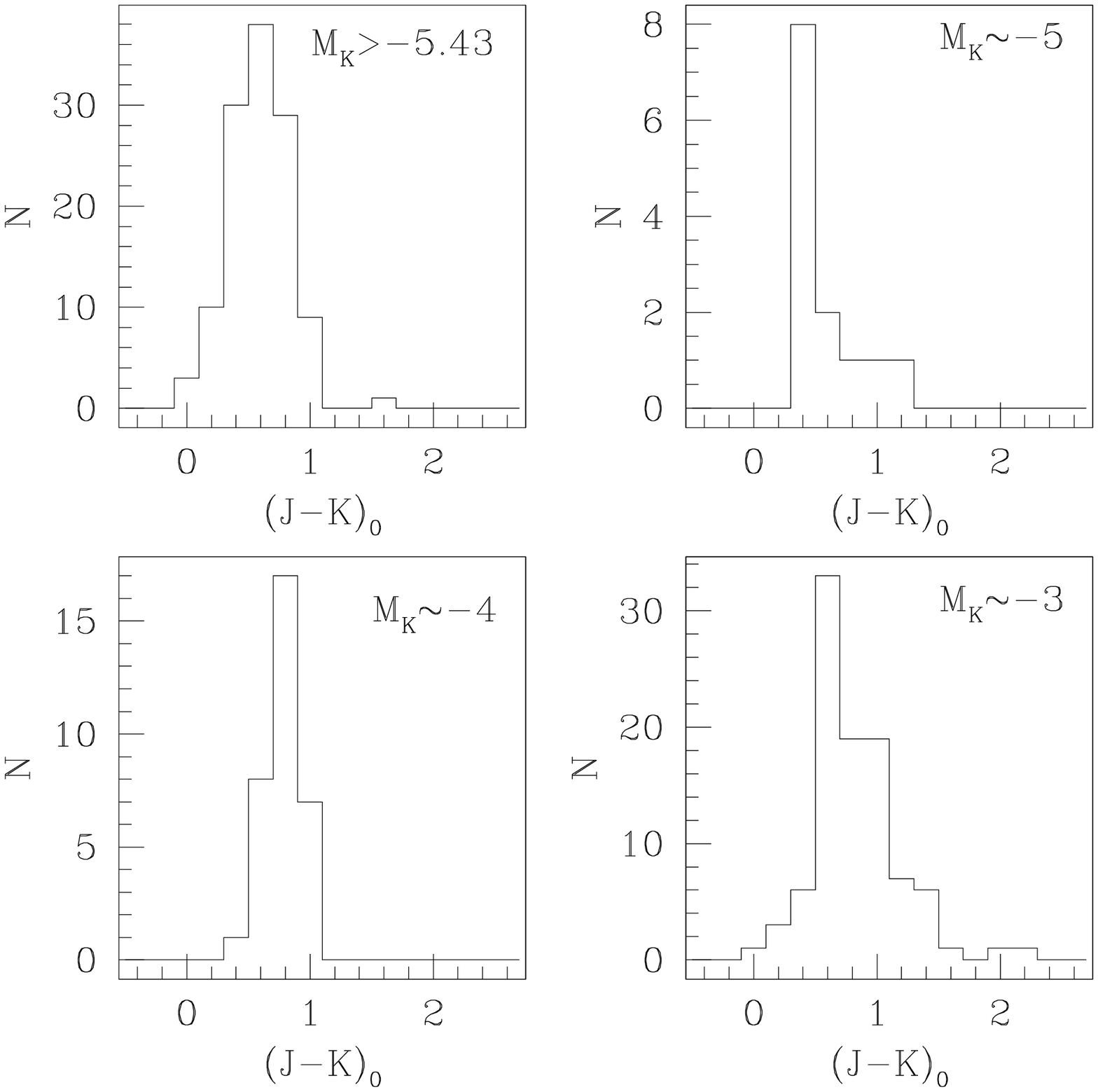}}
\end{minipage}
\caption{(Left)   Colour-magnitude  diagram   ($(J-K_{\mathrm  s})_0$,
$M_{K_{\mathrm s}}$)  of stars in Draco. Thick  slanted lines indicate
the  fiducial  RGBs, from  Ferraro  et  al.  (\cite{fer00}), of  three
globular clusters  (from left to right:  M15, M4 and  NGC 104).  Their
average metallicity is also indicated. Thick horizontal lines indicate
reference  magnitudes  around which  histograms  in  colour have  been
constructed with a bin size of $0.2$ mag (right).  These show the
distribution  of sources below  the TRGB  ($M_{K_{\mathrm s}}=-5.43$),
and  within $\pm0.05$ mag  of $M_{K_{\mathrm  s}}=-5$, $-4$  and $-3$,
respectively. The  binning is comparable to  the size of  the error in
the colour; it decreases with $1/\sqrt{N}$ in the histogram.}
\label{met}
\end{figure*}

Dereddening the  near-IR magnitudes and applying  the distance modulus
obtained in Sect. 4.1 it is possible to show that the mean metallicity
of Draco is [Fe/H]$=-1.95\pm1.26$ (where  the error is the variance of
the mean). Figure \ref{met} shows the distribution of Draco sources in
the  ($(J-K_{\mathrm   s})_0$,  $M_{K_{\mathrm  s}}$)  colour-magnitude
diagram.  These are  the  same  sources shown  in  Fig. \ref{jkk}  and
foreground stars  are still  included, which as  can be  easily judged
from their location, do not  affect the following discussion (at least
at  magnitudes  where  they  have been  confirmed).  Horizontal  lines
indicate  the location  of photometric  indices  that may  be used  to
determine [Fe/H]  (Valenti et al.  \cite{val}). In Fig.   \ref{met} we
show the  histogram of  all sources fainter  than the TRGB  (RGB stars
only)  and  three histograms  of  stars  within  $0.05$ mag  from  the
reference magnitudes $M_{K_{\mathrm s}}=-5$,  $-4$ and $-3$. The first
histogram   shows  a   fairly  symmetric   distribution   centred  at
$(J-K_{\mathrm s})_0=0.6$ with  FWHM$=0.2$. The other three histograms
peak at $(J-K_{\mathrm s})_0=0.4$,  $0.8$ and $0.6$, respectively, and
are not symmetric.  According to Valenti et al.  (\cite{val}) the peak
colours correspond to [Fe/H]$=-3.3$,  $-0.8$ and $-1.76$, for the three
histograms,  respectively. These  results do  not allow  us  to safely
determine  the  slope of  the  RGB,  especially  considering that  our
observations, though about $2$  mag deeper than 2MASS observations, do
not reach the  base of the RGB (about $7-8$ mag  below the TRGB). They
confirm,  however, that a  metallicity spread  within the  galaxy does
exist that amounts to about $1.26$ dex (variance of the mean) around a
mean value  of $-1.95$ dex.  Both the mean  and the spread  agree with
other  measurements  in  the  literature  (e.g.  Hartwick  \&  McClure
\cite{har},   Carney   \&   Seitzer   \cite{car},  Shetrone   et   al.
\cite{she01b}, Zinn \cite{zin78}, Winnick \cite{win}).

However three facts  need to be considered: (i)  there is a systematic
error in the $(J-K_{\mathrm s})_0$ colour due to the adjustment of the
photometry  to the  2MASS catalogue,  (ii)  metal poor  AGB stars  are
present    at    $-5>M_{K_{\mathrm    s}}>-6$    and    $(J-K_{\mathrm
s})_0\approx0.6$   (Sect.  4.3.2)     and  (iii)   the  foreground
contribution at faint magnitudes ($M_{K_{\mathrm s}}<-4$) has not been
determined.   Though the  statistical error  in the  colour  is about
$0.2$ mag, because of the larger error in the $J$ band compared to the
$K_{\mathrm s}$ band, it decreases  with the number of sources in each
bin as $1/\sqrt{N}$ and it  becomes on average $0.05$.  The systematic
error on the  colour amounts to about $0.09$  mag which corresponds to
an average  spread in the calculation  of [Fe/H] using  the Valenti et
al. (\cite{val}) indexes  of about $0.57$ dex. If  indeed sources with
$-5>M_{K_{\mathrm  s}}>-6$  and  $(J-K_{\mathrm  s})_0\approx0.6$  are
metal poor intermediate-age AGB stars they should be excluded from the
analysis of  the slope of  the RGB. This  is equivalent to  reduce the
peak in the histogram of  sources with $M_{K_{\mathrm s}}=-5$.  We
have  seen in  Sect. 3  that Galactic  foreground  with $M_{K_{\mathrm
s}}>-4$ stars  have ($0.2<J-K_{\mathrm s}<0.9$)  colours; their number
is  expected to  increase at  fainter magnitudes  and their  colour to
cover a larger  region if photometric errors become  also large (which
is  usually  the  case   approaching  the  sensitivity  limit  of  the
observations). Therefore we  conclude that the colour of  the RGB both
at $M_{K_{\mathrm s}}=-4$ and $M_{K_{\mathrm s}}=-3$ are uncertain.

 Another  way to estimate the metallicity  from the ($J-K_{\mathrm
s}$, $M_{K_{\mathrm  s}}$) colour-magnitude  diagram is to  compare it
with  fiducial   lines  of  Galactic  globular   clusters  with  known
metallicities    distance    and    reddening   (e.g.    Ferraro    et
al. \cite{fer00}). This  comparison with clusters M15, M4  and NGC 104
is also  shown in Fig. \ref{met}. These  clusters have [Fe/H]$=-2.12$,
$-1.19$  and  $-0.7$,  respectively.   The  thickness  of  the  rather
dispersed diagram of Draco sources is  fitted well by the RGB of these
clusters indicating an average [Fe/H]$=-1.34\pm0.72$ dex. A fit of the
TRGB is also clear indicating that  the majority of Draco stars are as
old as Galactic clusters.

\subsubsection{The population in three concentric areas}
Theory predicts that in two samples  of equal age the TRGB is brighter
in the sample with higher [Fe/H].  If the metallicity is the same, the
TRGB is fainter in the  older sample.  Using the isochrones by Girardi
et  al. (\cite{gir})  we have  calculated the  difference in  the TRGB
position  between populations  of $3$  and  $18$ Gyr  old and  between
populations    with   Z$=0.0004$    ([Fe/H]$=-1.7$)    and   Z=$0.001$
([Fe/H]$=-1.3$); these values were selected because we expect Draco to
be dominated  by an old  and metal-poor population. The  difference in
TRGB magnitude equals $0.244\pm0.003$  mag when the two samples differ
in age  and not  in metallicity and  to $0.284\pm0.003$ mag  when they
differ in metallicity but not in age. On average a difference of about
$0.26\pm0.3$  is expected  by  a variation  of  either metallicity  or
age. Between a  pair of regions (e.g. IN-MED)  we measure a difference
of about $0.5$ mag in the  location of the TRGB (Fig.  \ref{cmds}) and
this  can  only   be  explained  by  both  a   variation  in  age  and
metallicity. The  colour of the RGB  is also affected  by variations in
age  and metallicity:  at a  constant metallicity  younger  stars have
bluer RGBs  while at  a constant age  the RGB  of metal rich  stars is
redder than the  RGB of metal poor stars.  The average colour variation
in  either  age or  metallicity  expected  from  the same  populations
discussed above  amounts to  $0.10\pm0.03$ mag.  This  is of  the same
order as the variation in colour between a pair of regions and can thus
be  explained by  a  variation in  age  or metallicity  only  or by  a
combination of both.  However, the difference in colour  between the IN
and  the OUT  regions is  twice larger  and must  be due  to  both age
\textit{and} metallicity.   Another non-negligible contribution to
variations of the  TRGB location is of course  a variation in distance
within the galaxy.  Aparicio  et al. (\cite{apa}) suggested that Draco
extends about $14$ kpc along the  line of sight. This corresponds to a
variation of about $0.8$ mag in the TRGB position!

We now discuss  the apparent discontinuities in the  star counts shown
in the  lower part of  Fig. \ref{cmds}. The brighter  discontinuity in
the IN  region is equally well  described by Z$=0.0004$  and age $=18$
Gyr  as  by  Z$=0.001$  and  age  of  $3-18$  Gyr  while  the  fainter
discontinuity, if due to RGB  stars at the TRGB, is better represented
by  a slightly  more metal-rich  population of  comparable age.  It is
important,  however, to be  very careful  in deciding  which isochrone
best fits  the distribution of  sources because photometric  errors in
$M_{K_{\mathrm  s}}$ are  dominated  by the  systematic  error in  the
distance  modulus  ($0.15$ mag)  while  errors  in the  $(J-K_{\mathrm
s})_0$ are dominated by the error in the $J$ magnitude and affected by
a systematic error of about $0.09$ mag (Sect.  2). These errors do not
allow us to make firm  statements about an absolute age or metallicity
in a  given region but they  permit the analysis  of differences among
the  regions.  In  the  MED  region  stars  that  produce  the  bright
discontinuity are bluer  than stars with similar magnitudes  in the IN
region. This  means that in the  MED region they are  either AGB stars
with Z$=0.0004$  and age $= 1.6$  Gry or RGB stars  with Z$=0.001$ and
age $=  0.1$ Gyr. There is no  evidence of such a  young population of
giant stars  from optical studies  although a burst of  star formation
may have  happened $2-3$  Gyr ago (Aparicio  et al.   \cite{apa}). The
fainter discontinuity can be  explained by a population with Z$=0.001$
and age  $= 18$ Gyr while  a population with Z$=0.0004$  is bluer than
the  observed RGB  unless a  systematic photometric  shift  is needed,
i.e.   the  difference  is   due  to   measurement  errors.   The  two
discontinuities  in  the  OUT   region,  because  of  the  progressive
reddening  of the  RGB,  could be  due  to a  population of  increased
metallicity (that  could be as young  as a few Gyr).  The thickness of
the RGB around $M_{K_{\mathrm s}}=-4$  suggests the presence of also a
metal poor population,  but at these faint magnitudes  and blue colours
foreground  stars have  not been  studied and  their  distribution and
number may bias this conclusion.

Summarising: if most of the Draco population is as old as the Universe
($13.7\pm0.2$  Gyr; Bennett  et al.  \cite{ben}) then  its metallicity
increases with radius, which is  against the usual trend. Otherwise if
the spread in metallicity is not  very large then the variation in age
must  be large and  must increase  with radius  in agreement  with the
usual trend.   However, deeper and more  accurate photometric data
as well  as a treatment of  the orientation of the  galaxy will better
constrain these suggestions.

\section{Summary and Conclusions}
In this paper  we report observations in the  $I$, $J$ and $K_{\mathrm
s}$  band  covering  about  $40^\prime\times30^\prime$  of  the  Draco
galaxy. The photometry  has been calibrated to the  DENIS catalogue in
the $I$ band and to the 2MASS catalogue in the $J$ and $K_{\mathrm s}$
bands. The cross-identification  with catalogues of confirmed members,
non-members, carbon  and RGB  stars has helped  us to to  evaluate the
effect of  the foreground contribution  which has also  been discussed
using 2MASS  stars in  a comparable field  $0^{\circ}.4$ North  of the
centre.  Many  RGB stars have been  detected as well  as candidate AGB
stars.  The  TRGB location in the  $I$-band has been used  to derive a
distance        modulus         for        the        galaxy        of
$(m-M)_0=19.49\pm0.06(stat)\pm0.15(sys)$  in agreement with  values in
the literature and  the most recent determination from  RR Lyrae stars
(Bonanos et al. \cite{bon}).  The surface distribution of all detected
sources    traces   a   smooth    and   elliptical    structure   with
$\epsilon=0^{\circ}.27$ and  PA $\approx90^{\circ}$ also  in very good
agreement with previous measurements (Hodge \cite{hod}, Odenkirchen et
al.  \cite{ode}), and  suggests a  variation of  the  inclination with
radius.

 The  study of the  distribution of stars in  the colour-magnitude
diagram revealed  a population with  an average [Fe/H] of  $-1.95$ dex
(or $-1.34$ dex) with a spread  of $1.26$ dex (or at least $0.72$ dex)
using  the Valenti  et  al. (\cite{val})  photometric  indexes (or  by
comparison with the fiducial RGB  tracks of a few Galactic clusters). 
The  Girardi et al.  (\cite{gir}) isochrones  of two  populations with
[Fe/H]$=-1.3$ and  [Fe/H]$-1.7$ better represent  the colour-magnitude
diagrams of sources  located in three concentric regions.  They show a
non-negligible  intermediate-age population  even though  most  of the
stars are as old as the age of the Universe in agreement with Aparicio
et al.  (\cite{apa}) who  derived that some  $75-90$\% of  the stellar
population of Draco is older than $10$ Gyr.

\begin{acknowledgements}
We would like to thank the anonymous referee for comments and suggestions
that clearly improved the manuscript. 
This paper  makes use  of data  products from the  Two Micron  All Sky
Survey, which  is a joint  project of the University  of Massachusetts
and the  Infrared Processing and  Analysis centre/California Institute
of  Technology,   funded  by   the  National  Aeronautics   and  Space
Administration and the National Science Foundation.
\end{acknowledgements}


\begin{thebibliography}{}
\bibitem[1982]{aar} Aaronson,  M., Liebert,  J., Stocke, J.  1982, AJ,
254, 507
\bibitem[2001]{apa}
	Aparicio, A., Carrera, R., Mart\'{i}nez-Delgado, D. 2001, AJ, 122, 2524
\bibitem[1995]{arm}
	Armandroff, T.E., Olszewski, E.W., \& Pryor, C. 1995, AJ, 110, 2131
\bibitem[1986]{azz}
	Azzopardi, M., Lequeux, J., \& Westerlund, B.E. 1986, A\&A, 161, 232
\bibitem[1961]{bad}
	Baade, W. \& Swope, H.H. 1961, AJ, 66, 300
\bibitem[2002]{bel}
	Bellazzini, M., Ferraro, F.R., Origlia, L., et al. 2002, AJ, 124, 3222
\bibitem[2003]{ben}
	Bennett, C.L., Halpern, M., Hinshaw, G., et al. 2003, ApJS, 148, 1
\bibitem[1996]{sex}
	Bertin, E. \& Arnout, S. 1996, A\&AS, 117, 393
\bibitem[2004]{bon}
	Bonanos, A.Z., Stanek, K.Z., Szentgyorigvi, A.H., et al. 2004, AJ, 127, 861
\bibitem[1975]{can}
	Canterna, R. 1975, ApJ, 200, L63
\bibitem[1989]{cardi}
	Cardelli, J.A., Clayton, G.C., \& Mathis, J.S. 1989, ApJ, 345, 245
\bibitem[1986]{car}
	Carney, B.W., \& Seitzer, P. 1986, AJ, 92, 23
\bibitem[2005]{cio05}
	Cioni, M.-R.L., \& Habing, H.J. 2005, A\&A 429, 837
\bibitem[2004]{ciomes}
	Cioni, M.-R.L., Habing H.J., Loup, C., et al. 2004, The Messenger, 115, 22
\bibitem[2000b]{cio00}
	Cioni, M.-R.L., van der Marel, R.P., Loup, C., \& Habing,
H.J. 2000, A\&A, 359, 601
\bibitem[2000a]{cio00a}
	Cioni, M.-R.L, Loup, C., Habing, H.J., et al. 2000, A\&A, 144, 235 
\bibitem[2000]{fer00}
	Ferraro, F.R., Montegriffo, P., Origlia, L., \& Fusi Pecci,
F. 2000, ApJ, 119, 1282
\bibitem[1999]{fer99}
	Ferraro, F.R., Messineo, M., Fusi Pecci, F. et al. 1999, AJ, 118, 1738
\bibitem[1977]{deu}
	Deupree, R.G., \& Hodson, S.W. 1977, ApJ, 218, 654
\bibitem[2004]{dom}
	Dom\'{i}nguez, I., Abia, C., Straniero, O., et al. 2004, A\&A,
422, 1045
\bibitem[2003]{gal}
	Gallagher, J.S., Madsen, G.J., Reynolds, R.J., \& Grebel,
E.K. 2003, ApJ, 588, 326
\bibitem[2000]{gir}
	Girardi, L., Bressan, A., Bertelli, G., \& Chiosi, C. 2000,
A\&AS, 141, 371
\bibitem[1998]{gri}
	Grillmair, C.J., Mould, J.R., Holtzman, J.A., et al. 1998, AJ, 115, 144
\bibitem[1996]{har}
	Hargreaves, J.C., Gilmore, G., Irwin, M.J., Carter, D. 1996,
	MNRAS, 282, 305
\bibitem[1974]{hart}
	Hartwick, F.D.A., \& McClure R. 1974, ApJ, 193, 321 
\bibitem[1964]{hod}
	Hodge, P.W. 1964, AJ, 69, 853
\bibitem[1998]{hunt}
	Hunt, L.K., Mannucci, F., Testi, F. et al. 1998, AJ, 115, 2594
\bibitem[2002]{iku}
	Ikuta, C., \& Arimoto, N. 2002, A\&A, 391, 55
\bibitem[1995]{irw}
	Irwin, M., \& Hatzidimitriou, D. 1995, MNRAS, 277, 1354
\bibitem[1992]{land}
	Landolt, A.U. 1992, AJ, 104, 340
\bibitem[1993]{lee}
	Lee, M.G., Freedman, W.L., \& Madore, B.F. 1993, ApJ, 417, 553
\bibitem[1992]{len}
	Lehnert, M.D., Bell, R.A., Hesser, J.E., \& Oke, J.B. 1992, ApJ, 395, 466
\bibitem[1995]{mad}
	Madore, B.F., \& Freedman, W.L. 1995, AJ, 109, 1645
\bibitem[1998]{mat}
	Mateo, M. 1998, ARA\&A, 36, 435
\bibitem[1991]{mun}
	Munari, U. 1991, A\&A, 251, 103 
\bibitem[1985]{nem}
	Nemec, J.M. 1985, AJ, 90, 204
\bibitem[2000]{nik}
	Nikolaev, S., \& Weinberg, M.D. 2000, ApJ, 542, 804
\bibitem[2001]{ode}
	Odenkirchen, M., Grebel, E.K., Harbeck, D., et al. 2001, AJ, 122, 2538 
\bibitem[1996]{ols}
	Olszewski, E.W., Prvor, C., \& Armandroff, T.E. 1996, AJ, 111, 750
\bibitem[1995]{ols95}
	Olszewski, E.W., Aaronson, M., \& Hill, J.M. 1995, AJ, 110, 2120
\bibitem[2002]{pia}
	Piatek, S., Pryor, C., Armandroff, T.E., et al. 2002, AJ, 123, 2511
\bibitem[2003]{rav}
	Rave, H.A., Zaho, C., Newberg, H.J., et al. 2003, ApJS, 145, 245
\bibitem[1997]{sal}
	Salaris, M., \& Cassisi, S. 1997, MNRAS, 289, 406
\bibitem[1998]{she98}
	Shetrone, M.D., Bolte, M., \& Stetson, P.B. 1998, AJ, 115, 1888
\bibitem[2001a]{she01a}
	Shetrone, M.D., C\^{o}t\'{e} P., \& Stetson, P.B. 2001, PASP, 113, 1122
\bibitem[2001b]{she01b}
	Shetrone, M.D., C\^{o}t\'{e}, P., \& Sargent, W.L.W. 2001, AJ, 548, 592
\bibitem[1984]{smi}
	Smith, G.H. 1984, AJ, 89, 801
\bibitem[1979a]{ste79a}
	Stetson, P.B. 1979, AJ, 84, 1149
\bibitem[1979b]{ste79b}
	Stetson, P.B. 1979, AJ, 84, 1167
\bibitem[1980]{ste80}
	Stetson, P.B. 1980, AJ, 85, 398
\bibitem[1984]{ste84}
	Stetson, P.B. 1984, PASP, 96, 128
\bibitem[1985]{ste85}
	Stetson, P.B., McClure, R.D., \& van den Berg, D.A. 1985, PASP, 97, 908
\bibitem[2004]{val}
	Valenti, E., Ferraro, F.R., \& Origlia, L. 2004, MNRAS, 351, 1204
\bibitem[2000]{vdb}
	van den Berg, S. 2000, In: The Galaxies of the Local Group,
Cambridge University Press
\bibitem[2004]{wil}
	Wilkinson, M.I., Kleyna, J.T., Evans, N., et al. 2004, ApJ, 611, L21
\bibitem[2003]{win}
	Winnick, R. 2003, Ph.D. Thesis, Yale University
\bibitem[1999]{you}
    Young, L.M. 1999, AJ, 117, 1758
\bibitem[1978]{zin78}
	Zinn, R. 1978, ApJ, 225, 790
	
\end{thebibliography}
\end{document}